\documentstyle[12pt]{article}
\input epsf
\begin{document}
\begin{titlepage}

\begin{tabular}{l}
January, 1995                   
\end{tabular}
    \hfill
\begin{tabular}{l}
MSU-HEP-41027 \\
CTEQ-407 \\
\end{tabular}

\vspace{2cm}

\begin{center}

\renewcommand{\thefootnote}{\fnsymbol{footnote}}

{\LARGE A Global QCD Study of Direct Photon Production\footnote[1]{
This work was supported in part by DOE, NSF, and TNRLC.}}

\renewcommand{\thefootnote}{\arabic{footnote}}

\vspace{1.25cm}


{\large J.~Huston,$^d$ E.~Kovacs,$^{b,\dag}$ S.~Kuhlmann,$^a$ H.~L.~Lai,$^d$
J.~F.~Owens,$^c$ W.~K.~Tung$^d$}

\vspace{1.25cm}

$^a$Argonne National Laboratory,
$^b$Fermi National Accelerator Laboratory, \\
$^c$Florida State University,
$^d$Michigan State University, \\
$^{\dag}$Visitor

\end{center}
\vfill

\begin{abstract}
A global QCD analysis of the direct photon production process from both fixed
target and collider experiments is presented. These data sets now completely
cover the parton $x$ range from 0.01 to 0.6, thereby providing a stringent test
of perturbative QCD and parton distributions. Previous detailed studies of
direct photons emphasized fixed target data. We find most data sets have a
steeper $p_t$ distribution than the QCD prediction. Neither global fits
with new parton distributions nor improved photon fragmentation
functions can resolve this problem since the deviation occurs at different $x$
values for experiments at different energies. A more likely explanation is the
need for additional broadening of the $k_t$ of the initial state partons.
The magnitude and the
possible physical origin of this effect are investigated and
discussed.
\end{abstract}

\vfill
\newpage
\end{titlepage}

\section{Introduction\label{sec:intro}}

Direct photon production in hadron collisions is an important process to
study in perturbative Quantum Chromodynamics (QCD) due to the clean
measurement of photons, in contrast to jet physics with the jet definition
uncertainties. In addition, sensitivity
to the initial state gluon in the important Compton scattering mechanism
provides a valuable opportunity to directly measure the gluon content inside
the proton. These features have been utilized in studies \cite{abfow} of
earlier direct photon experiments in the context of next-to-leading-order
(NLO) ($\alpha \alpha _s^2$) QCD calculations of Aurenche {\it et al.}~\cite
{aur}, and in the global analyses of parton distributions.\cite{mrsb,mrsd,cteq}
The potential usefulness of the direct photon experiments demonstrated in
these works has yet to be fully realized due to the limited range
of the fixed target experiments as well as remaining theoretical
uncertainties related to scale-dependence and fragmentation processes.

In hadron colliders, earlier data from the ISR and $Sp\bar pS$ colliders
have been
recently augmented by high statistics data from the Fermilab TEVATRON. The $%
x $-range covered by these collider and fixed-target experiments now spans
the entire interval ($0.01<x<0.6$). The combined range and improved accuracy
of these data, as well as recent theoretical developments, provide a good
opportunity to make a global comparison between theory and experiment. We
report here such a study. The next two sections examine existing work on
fixed-target and collider data and treat them in a unified context. A
pattern of deviation of the observed $x_t$ distribution from existing QCD
expectations is highlighted. This is followed by a new global analysis of all
existing data. We show that it is difficult to account for the observed
pattern by new sets of parton (mostly gluon) distributions because the
deviations occur at different $x$ values for different experiments. Various
alternative origins of the observed pattern and possible related phenomena
in other processes are discussed.

\section{Fixed Target Results\label{sec:fixedT}}

\begin{figure}[htbp]
\epsfxsize=\hsize \epsffile{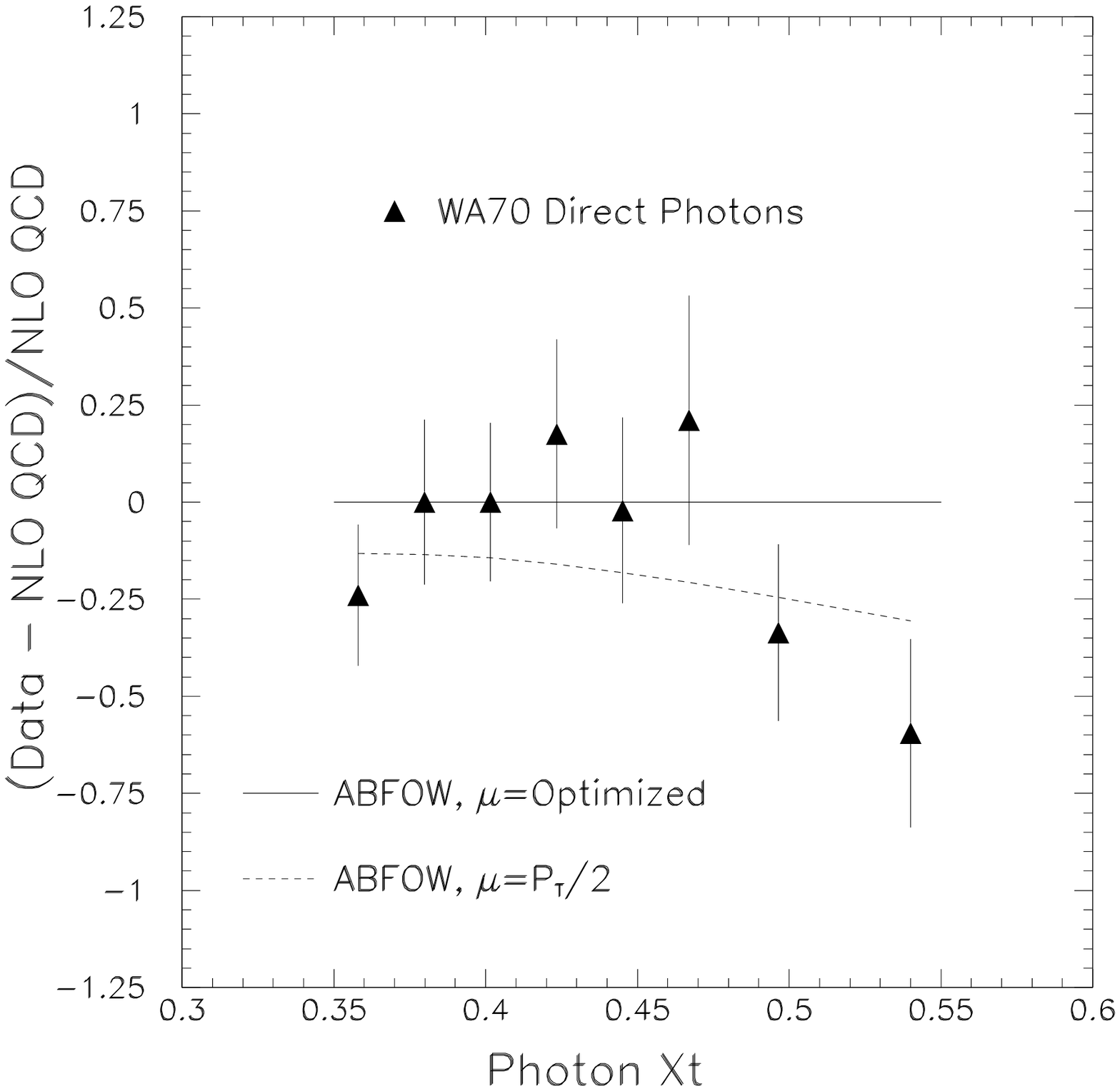}
	\caption{Direct photon production data from WA70 is compared to
the NLO QCD fit of ABFOW using optimized scales.
The fractional difference is shown in order to display
details of the comparison.
(see text)}
	\label{fig:wa70plot}
\end{figure}

Previous parton distribution analyses incorporating direct photon data have
focused on fixed target experiments, particularly the WA70 experiment (pp
collisions at $\sqrt{s}=23$ GeV~\cite{wa70}) from CERN. The ABFOW~\cite
{abfow} analysis performed fits on DIS data from the BCDMS experiment and used
WA70 data as constraints to determine the gluon distribution, then
checked many other direct photon experiments with the
resulting parton distributions. The WA70
experimental normalization was fixed at unity.\footnote{%
Since the Compton cross-section is proportional to $\alpha _s(\mu ,\Lambda
)\cdot G(x,\mu )$, the normalization factor directly affects the
determination of $\Lambda $ and $G(x,\mu )$.\label{fn:normalization}} The
MRS analyses~\cite{mrsb,mrsd}~used a similar procedure but included more DIS
data and added Drell-Yan results as constraints. Both of these groups used
the ``optimized scale'' scheme \cite{opt} in choosing the renormalization
and factorization scales. Fig.~\ref{fig:wa70plot} shows the comparison of
WA70 data with the ABFOW fit in the form (Data~-~NLO~QCD)/NLO~QCD as
a function of
photon $x_t$ ($2p_t/\sqrt{s}$). The production is fairly central so that
photon $x_t$ is approximately the parton $x$ value probed. One sees that the
fit is satisfactory. To see the scale-dependence of the NLO theory, one can
compare the results obtained from two choices of scale, for example
optimized and $p_t/2$. The latter is
normalized to the former (as are the data points) and
shown in the plot as the dashed line.
The optimization procedure results in
higher cross-sections since it yields scales smaller than $p_t/2$ for this
kinematic region.

\begin{figure}[htbp]
\epsfxsize=\hsize \epsffile{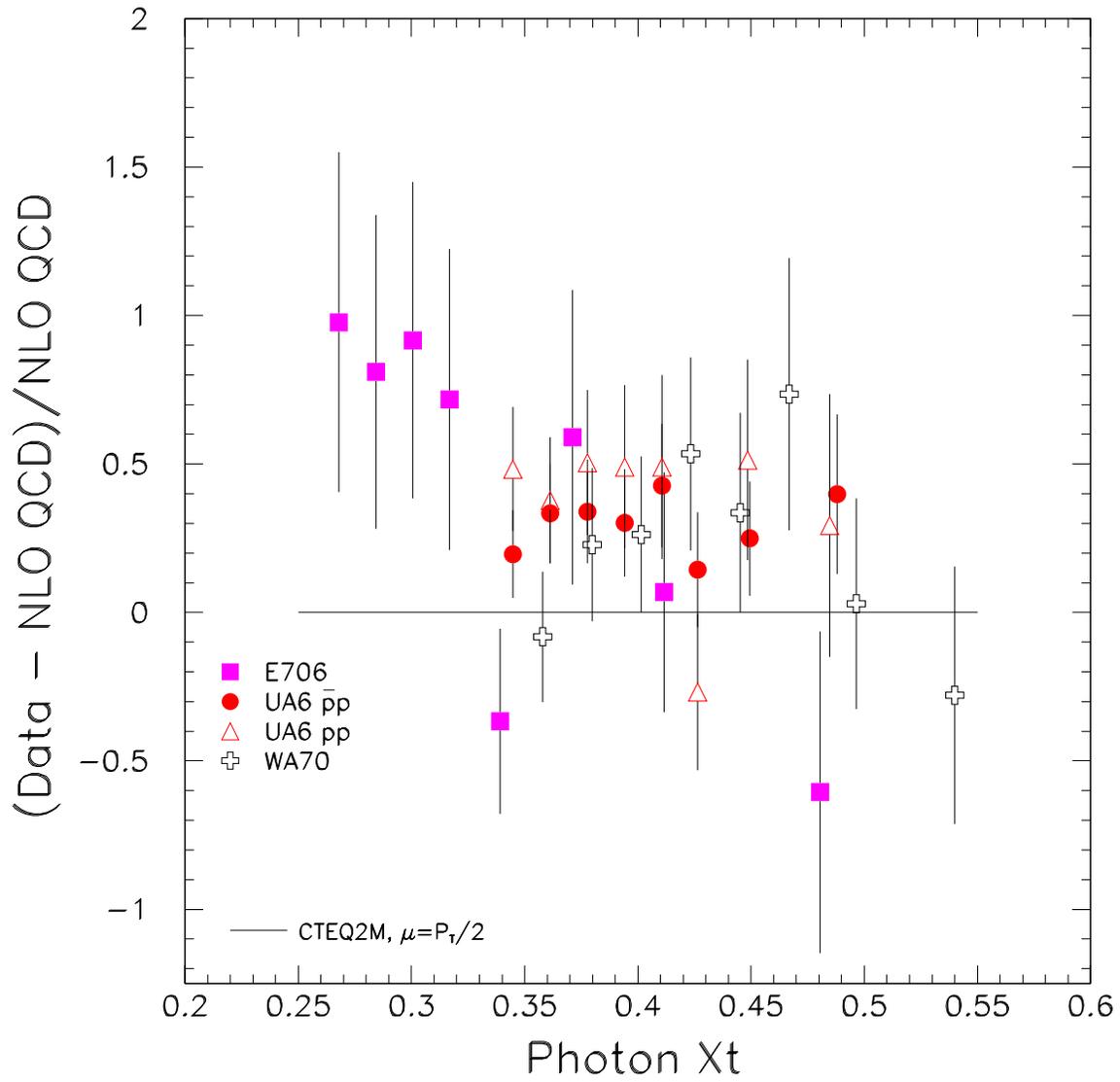}
	\caption{Fixed target direct photon
experiments WA70, UA6, and E706 are compared to a NLO QCD results. Fractional
differences
between data points and the CTEQ2M fit are shown. (see text)}
	\label{fig:ftdpComp}
\end{figure}

In the CTEQ2~global analysis\cite{cteq}, in addition to WA70, data from the
fixed target experiments UA6 ($pp\ $collisions at $\sqrt{s}=24$ GeV~\cite
{ua6-1}) and E706 (pBe collisions\footnote{The nuclear target dependence
has been studied by E706 and is consistent with A$^{1.0}$ for
direct photon production\cite{roser}.} at $\sqrt{s}=30.7$ GeV~\cite{e706}) were
used. These data were treated on the same footing as the DIS and Drell-Yan
data in an overall global fit. The uncertainties on experimental
normalization and theoretical scale-dependence are both taken into account
in the analysis procedure (see Sec.\ref{sec:Global} and Ref.\cite{cteq} for
details). In Fig.~\ref{fig:ftdpComp} data from these three experiments are
compared to the CTEQ2M fit.  In this figure the more recent data
from UA6 (pp and $\bar p p$ collisions~\cite{ua6-2}), which agree with
their earlier data, are shown. The absolute normalizations
of all data and theory are used in this figure.
These fixed-target experiments cover similar $x_t$ ranges (with E706 extending
somewhat lower) and are roughly consistent with each other. We note that the
E706 data show a deviation from NLO QCD on this comparison plot
which was not evident in previous comparisons with WA70 data alone. This
pattern is also seen with other current parton distribution sets such as
MRSD0~\cite{mrsd}.

\section{Collider Results and Calculations\label{sec:colliderR}}

The first direct photon data from hadron colliders came from the ISR
pp collider\cite{r806}, followed by the $Sp\bar pS$ collider~\cite{ua2}.
These data were found to be in
qualitative agreement with NLO theory in Ref.\cite{abfow}. However, the
1989 data from the TEVATRON $\bar pp$ collider experiment CDF~\cite{cdf89},
extended into a new $x$~region~($0.01$),
and showed a clear deviation ($1.5-2$ times higher)
from the NLO QCD predictions
below $p_t\sim 20$ GeV. Improved parton distribution sets such as CTEQ1~\cite
{cteq} and MRSD0~\cite{mrsd}, using newer small-$x$ deep inelastic scattering
data, removed approximately one-half of
this discrepancy,  but the remaining difference
persisted. This effect has recently been confirmed with much higher
statistics data~from CDF \cite{cdf}. The obvious source of uncertainty due
to choice of scale cannot be responsible for the discrepancy since it
produces small normalization shifts with no change in shape~\cite{cdf89}. In
addition, the optimized scales scheme, favored in fixed-target fits, was
shown to make the disagreement between data and NLO QCD worse~\cite{cdf89}.
Thus, much attention has been directed to the proper theoretical treatment of
{\em isolated} photon cross-sections (which the collider experiments
measure) and the related issue of the photon fragmentation
function.\cite{qiu,Chiapetta,owens,dort} This is
relevant since the collider
kinematic range is sensitive to the {\it bremsstrahlung} process, where a
photon is radiated off an initial or final state quark.

Analytic NLO calculations mentioned earlier \cite{aur} include the {\it %
bremsstrahlung} process, and use a leading-order (LO) fragmentation function
in the region where the photon is emitted collinear to the quark. The
implementation of the isolation cut in these calculations involves some
approximations, and the (minor) effects of soft gluons inside the isolation
cone are ignored. Alternatively, a Monte Carlo (MC) program for calculating
NLO direct photon production (Baer {\it et al.}~\cite{owens}) has proved to
be very useful, particularly for colliders, since it provides a natural way
to implement isolation cuts and the soft gluon effects.
The MC calculation yields fully-inclusive
cross sections within a few \% of the analytic method. For calculations
reported below, we compute the ratio of isolated-photon to total-inclusive
cross sections using the MC method.  We then use this ratio to put the collider
and fixed-target data sets on the same footing in order to make a global
comparison of data and the analytic calculations, and to
perform new global fits for
parton distributions. For the global
fits, this procedure is also necessary since only~the analytic form of
the inclusive cross-section is efficient enough for the required repetitive
calculations.

\begin{figure}[htbp]
\epsfxsize=\hsize \epsffile{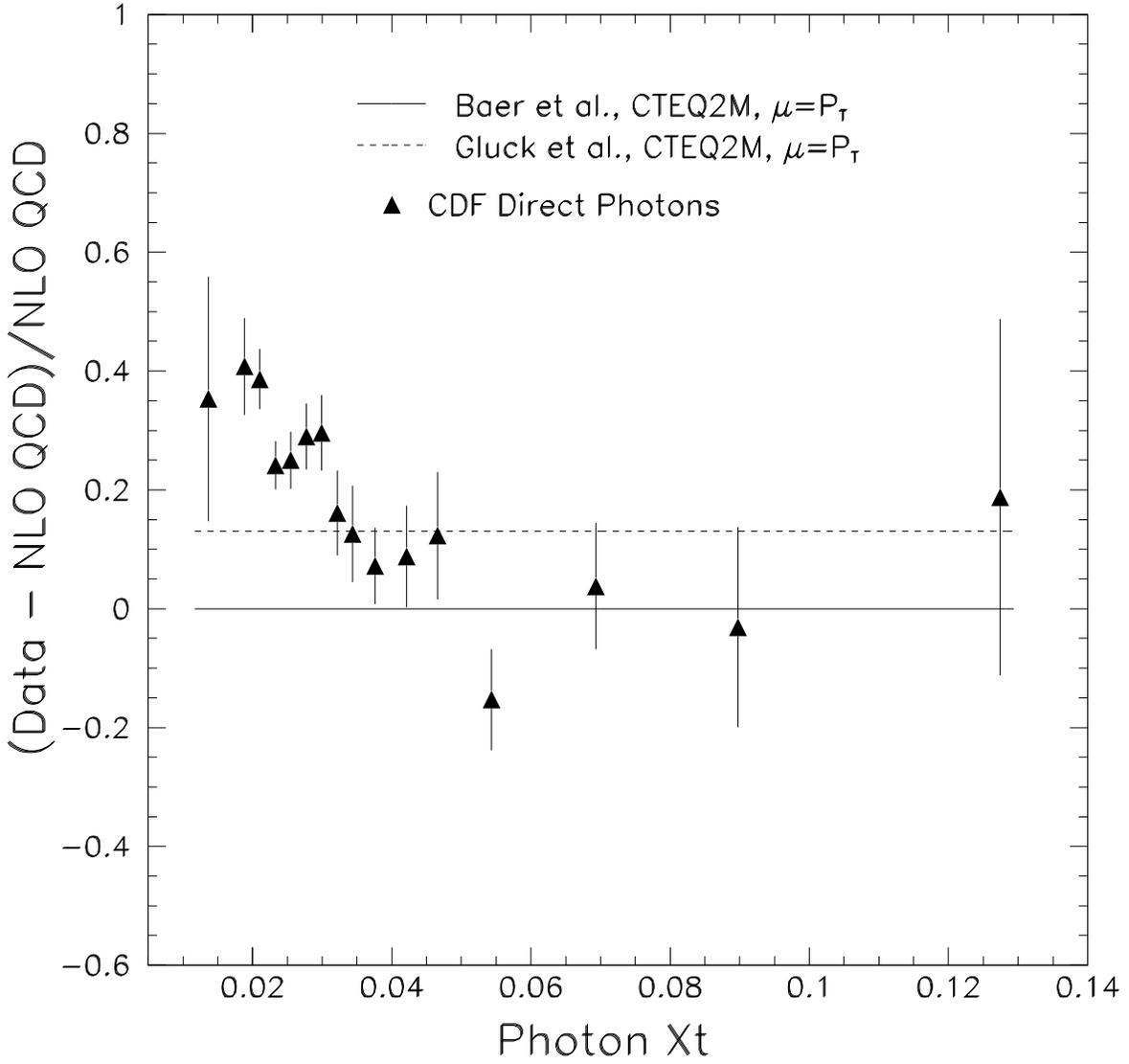}
	\caption{Direct photon data from the CDF experiment
is compared to NLO QCD predictions. (see text)}
	\label{fig:dortxt}
\end{figure}

The analytic and MC programs mentioned above both use a LO fragmentation
function convoluted with the LO two-jet production diagrams. There is now an
analytic calculation from Gluck {\it et al.}~\cite{dort} which uses a NLO
parameterization of the photon fragmentation function convoluted with LO
two-jet production, and NLO jet production convoluted with the LO photon
fragmentation function. In Fig.~\ref{fig:dortxt}, we compare this
calculation, as well as that of Baer {\it et al.}~\cite{owens}, to the new
CDF data. The two curves are evaluated with the same parton distributions
(CTEQ2M) and scales ($\mu =p_t$). The Gluck {\it et al.} calculation is 13\%
higher than Baer {\it et al.} uniformly in $p_t$. Thus the observed shape of
the CDF data does not agree with either calculation. We note that the
calculation of Ref. \cite{dort} used GRV~\cite{grv} parton distributions and
a scale of $\mu =p_t/2$ (except for the fragmentation scale where $\mu =$%
(Cone-size)\thinspace $*p_t=0.7*p_t$ was used). The scale change leads to a
10\% normalization shift upward, while the different parton distributions
provide a 5\% shape change at photon $p_t=16$ GeV.  With these parameters
the total fragmentation contribution to the isolated cross section
was estimated in Ref.~\cite{dort} to be $\approx 10\%$, hence any modification
of this contribution is unlikely to explain the $\approx 30\%$
difference in shape between the CDF data and NLO QCD calculation.
In addition, the strength of the CDF data
is in their small (5\%) systematic uncertainty on the measured
shape versus $p_t$,  hence this
new calculation is still insufficient to explain the data completely.

\section{Global Comparison of Data and NLO QCD\label{sec:Global}}

\begin{figure}[htbp]
\epsfxsize=\hsize \epsffile{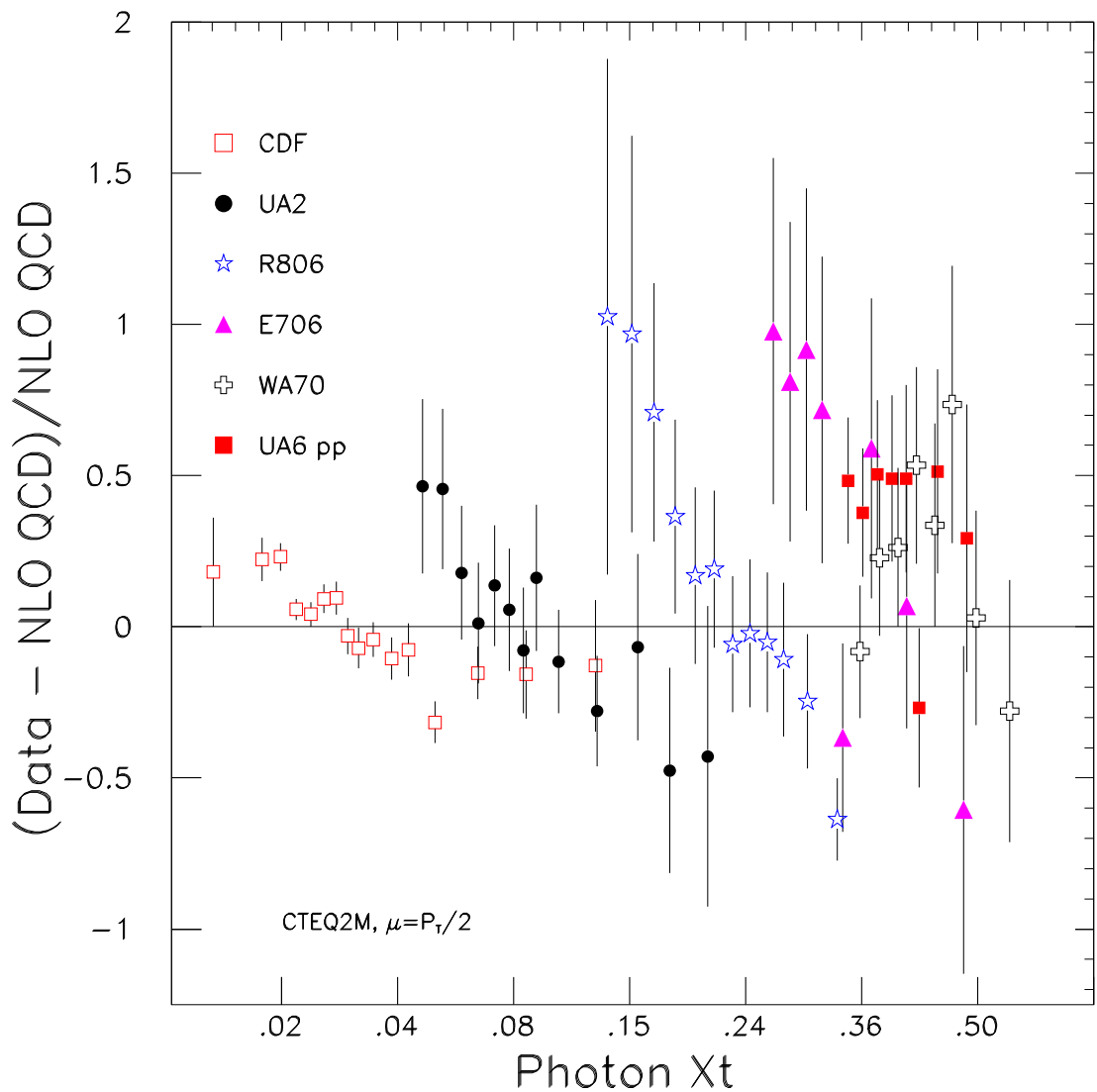}
	\caption{Compilation of direct photon experiments compared to
NLO QCD predictions using CTEQ2M parton distributions. (see text)}
	\label{fig:allct22}
\end{figure}

The fixed-target and collider experiments described above together cover the
photon $x_t$ range from $0.01\ $to\ $0.6$. It is natural to examine all the
data at once in the framework of NLO QCD and to utilize the combined power of
this rich collection of data in a new
global NLO QCD+parton distribution analysis. In Fig.\ref
{fig:allct22} we show the compilation of data sets from colliders and
fixed target experiments
compared to NLO theory (using CTEQ2M parton
distributions and $\mu =p_t/2$). We see for the first time the impressive
full coverage of the entire $0.01<x<0.6$ range.
We also see, however, that
most of the data sets display a steeper dependence on $x_t$ than
is predicted by NLO QCD (horizontal line).
We note that since
both theory and experiment have inherent normalization uncertainties, it is
difficult to discern an ``excess'' of photons at low $p_t$ from a
``deficit'' at high $p_t$.  However, the difference in slope between most
data sets and the theory is quite apparent, and does not depend strongly
on the experiment's $x_t$ range.

All previous global analyses used only fixed-target direct photon data in
the fit. One obvious study is to perform a new
global analysis including all the photon data in order to see to what extent
new parton distributions, especially the gluon distribution which is not
well constrained so far, could reduce the observed discrepancy.
In this study, the most recent direct photon data from CDF \cite{cdf} and
UA6 (both $pp$ and $\bar pp$) \cite{ua6-2} are included together with UA2,
R806, E706, WA70 and the standard DIS and Drell-Yan data sets used in the
CTEQ2 global analysis~\cite{cteq}. The systematic errors for each experiment
were separated into a normalization error (typically 10-15\%) and
point-to-point systematic errors that were added in quadrature with the
statistical errors. The normalization for each experiment was allowed to
vary from unity, constrained by the appropriate experimental errors as part
of the $\chi ^2$ minimization procedure. The theoretical scale uncertainty
was taken into account by assigning a ``theoretical error''
calculated using the difference obtained with scale choices of $p_t/2$ and $%
2p_t$. The scale $\mu $ was then allowed to vary during the minimization
process as an overall parameter with the associated error.

\begin{figure}[htbp]
\epsfxsize=\hsize \epsffile{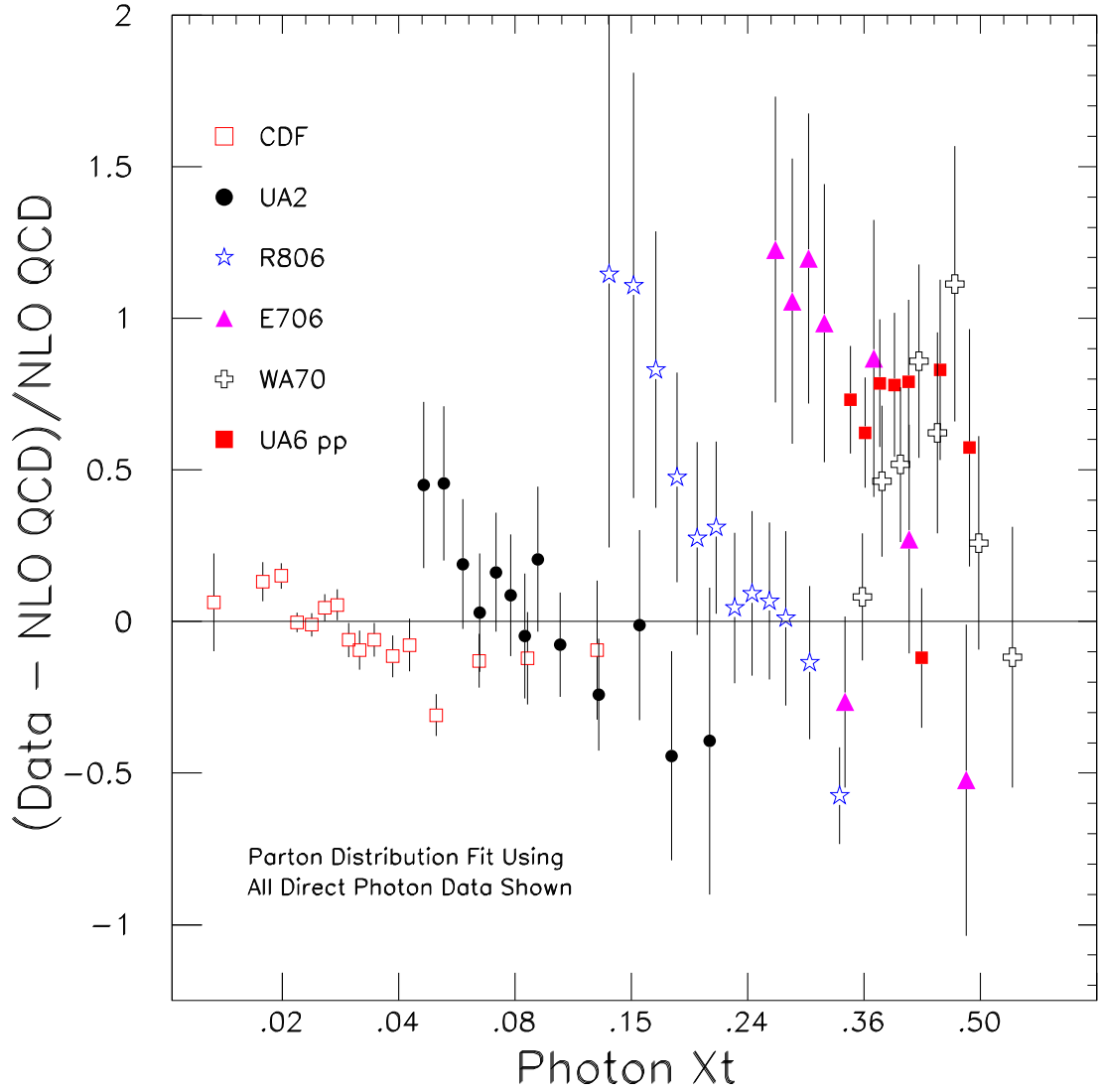}
	\caption{Compilation of direct photon experiments compared to
a NLO QCD prediction using parton distributions fit using all the
data shown. (see text)}
	\label{fig:allgt05}
\end{figure}

Comparison of results from such a representative fit
to the data points is shown in Fig.~\ref{fig:allgt05}.
The experiments are shown with their absolute normalization;
in the fit they are renormalized downward by $\approx 20\%$.
Qualitatively, this looks similar to Fig.~\ref{fig:allct22}. However,
the CDF data do come closer to the theory, while the fixed target data in
particular move farther away.  Clearly, the small statistical and systematic
errors of the CDF experiment cause this data set to dominate the fit, hence
increasing the gluon density $G(x,Q)$ in its $x$-range. This increase is
compensated by a decrease in the gluon density in the large-$x$ region due
to the momentum sum rule
constraint.\footnote{%
The momentum fraction carried by the gluon is constrained
by the momentum sum rule to the range
0.42-0.43 in most current global parton analyses,
since the total momentum fraction carried by quarks is very well determined
by recent precision DIS experiments. We also note that the $x$ range of the CDF
data (0.01-0.1) carries a large percentage of the gluon momentum, given the
general shape of the gluon distribution.}
This demonstrates, as expected, that it
is difficult to improve the differing slopes by new parton
distributions.  Furthermore, we note that it is also difficult to attribute
these differing slopes to the photon
fragmentation function since the ISR and fixed target data are
insensitive to fragmentation.

\section{Discussion of Global Analysis Results}

What could cause an effect seen in most experiments at low $p_t$
irrespective of their $x$ range? One possibility is that of ``$k_t
$ smearing'', since any uniform smearing on a steeply falling $p_t$
distribution enhances the low $p_t$ end of the spectrum more than the
high $p_t$ end. The question is therefore: whether the NLO
QCD calculation (involving at most one non-collinear initial
state gluon) embodies sufficient $k_t$ broadening of the
initial state partons, or
whether additional effects due to initial state
multi-gluon radiation and non-perturbative physics (i.e. ``intrinsic $k_t$'')
also need to be included?

In order to explore this problem in a simplistic fashion, the NLO QCD
prediction for the cross section was convoluted with gaussian
functions of transverse momentum.  The effect of such a
convolution is to induce corrections to the cross section
that behave as $\sim \sigma^2/{p_t}^2$, where $p_t$ is the
photon $p_t$ and $\sigma$ is the width of the
gaussian, representing the additional $k_t$ in the system.
The values of $\sigma$ needed to reproduce
approximately the collider data sets varied from
$\approx $2-3 GeV for R806, to $\approx $3-4 GeV
for UA2 and CDF.
The corrections of course also depend on the local slope
of the NLO QCD cross sections, therefore the
fixed target experiments are extremely sensitive
to $\sigma$ due to their rapidly falling $p_t$ spectra.
A width of $\approx $1-2 GeV for E706, and slightly
smaller than 1 GeV for WA70 and UA6 gave a reasonable overall
description of all the data sets.
This exercise is, of course, completely ad hoc: it merely
serves the purpose of indicating the approximate magnitudes of the smearing
necessary to bridge the gap between the
NLO QCD calculation and experiment. The energy
dependence of the required smearing width reveals that this possible
effect is probably not of
a simple ``intrinsic'' type associated with the confinement size of the
hadron, but more likely due to multiple gluon emissions.\footnote{%
In the previous analysis in ref.~\cite{abfow} mentioned earlier,
there was no conclusion of
the need for an additional contribution from initial state $k_t$.
However, some
elements were present in that analysis that might have masked this effect:
(i) they emphasized the WA70 data,
which are qualitatively consistent with NLO QCD
without additional smearing;
(ii) the low $p_t$ data points from the ISR (where the
deviation is most apparent) were removed in the analysis because the
scale optimization failed;
and (iii) the discrepency between the
low $p_t$ data points of UA1 \& UA2 and theory was attributed to
uncertainties of photon fragmentation contribution and isolation cuts at the
time (recent progress reviewed in Sec.\ref{sec:colliderR} has
considerably reduced these uncertainties).}

Smearing due to multiple gluon emissions is incorporated in many QCD shower
MC programs. As another exercise, the QCD MC program PYTHIA~\cite{pythia}
was utilized to generate the direct photon cross section for the CDF
experiment used in our analysis. Within PYTHIA, there is the option to turn
on/off initial state gluon radiation. We found that the ratios of cross
sections (initial state radiation on/ initial state radiation off) show the
same general pattern for the CDF data as seen in Figures~%
\ref{fig:allct22} and \ref{fig:allgt05}. The MC program uses only LO
hard matrix elements, and
the relation of $k_t$ smearing from the multiple gluons
in this calculation to the fixed order NLO cross-section is not clear. However,
it is obvious that physical effects due to multiple radiation have the
general characteristics of the observed cross-section.

Theoretically, multiple gluon emission is the main physics effect treated
systematically in the ``$p_t$ resummation'' formalism of vector-boson
production (continuum Drell-Yan, W, and Z production) \cite{CCS}. Although,
in principle, $p_t$ resummation is not needed for inclusive direct
photon production (because it is not strictly speaking a ``multiple large
scale'' problem), it is intriguing to note that the resummed cross-section
formula of Collins {\it et al. }~\cite{CCS} requires a ``non-perturbative''
broadening factor which bridges the $p_t$ physics at the confinement and
perturbative scales, and this broadening is inherently energy dependent. It is
well worth exploring, both theoretically and phenomenologically, any
possible relations between this result and the question at hand.

\begin{figure}[htbp]
\epsfxsize=\hsize \epsffile{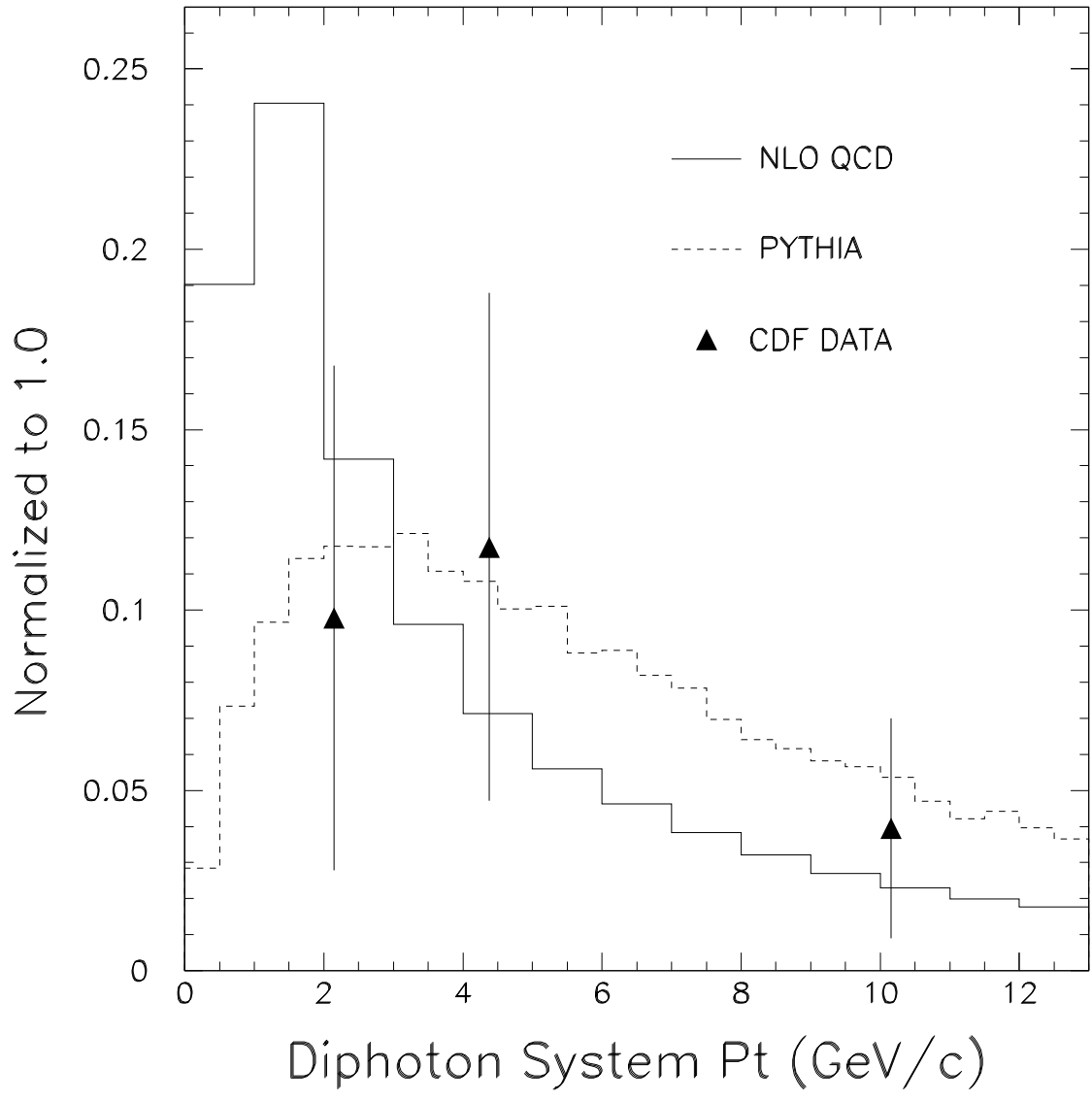}
	\caption{Diphoton system $p_t$ distribution for PYTHIA, NLO QCD, and
CDF data. (see text)}
	\label{fig:ctdipho}
\end{figure}

{}From an experimental point of view, one can investigate $k_t$ effects
directly by examining the $p_t$ imbalance of diphoton states. WA70
\cite{wa70dipho} compared the distribution of diphoton system $p_t$ with
that predicted by NLO QCD smeared with a gaussian $k_t$ broadening.
They found that the typical width of
 0.34 GeV intrinsic parton transverse momentum was
insufficient to explain their data, while an additional broadening
 of 0.91 GeV gave a much
improved fit.  This extra broadening for diphotons is close to the amount
that gave a reasonable overall description of the fixed target
single photon measurements
mentioned earlier. A similar situation exists for CDF diphotons~\cite
{cdfdipho}. Fig.~\ref{fig:ctdipho} shows the diphoton system $p_t$ from
the data, NLO QCD and PYTHIA. The small
statistics of the data do not provide strong discrimination at this time,
but the
important point is that the distribution from PYTHIA is significantly
broader than NLO QCD. Once again if the NLO QCD prediction is smeared with a
gaussian, a width of $\approx $3 GeV is needed to reproduce PYTHIA, and this
is close to what was needed to bring the CDF {\it single} photons in agreement
with NLO QCD.  The agreement in the magnitude of the broadening
necessary for the diphotons and the single photons may be fortuitous,
since the production processes are different.  But the important point is
that the diphotons indicate a potential problem with $k_t$, complementing
the inclusive photon trends.

\section{Conclusions and Future Prospects}

In conclusion we have performed a global study of direct photon
experiments spanning a wide range of parton $x$ values. We see a pattern of
deviation between the measured slopes versus photon $p_t$ and
the corresponding NLO QCD calculations.
The deviations occur in several different experiments in
different parton $x$ ranges with widely varying sensitivities
to photon fragmentation functions and theory scales.
Thus it is difficult to invoke new
parton distribution functions or a different treatment of the fragmentation
process to explain all the data. This is verified by a new global parton
distribution fit using all the direct photon data described in this paper.
One possibility is the NLO QCD theory needs
to be supplemented by additional
$k_t$ broadening from non-perturbative physics and/or initial
state gluon radiation.
A similar phenomenon has been seen in studies of
the relative $p_t$ of diphoton final states and this premise has been
considered for some time~\cite{workshop}.
We note that the same effect should also appear in low $p_t$ measurements
of other processes such as dijet and heavy flavor production.
Low $p_t$ dijet production typically has substantial systematic
uncertainties~\cite{dijet}, but
the effects of $k_t$ {\it have} been extensively studied in
fixed target heavy flavor production~\cite{mlm}, also with the
conclusion for the need for additional $k_t$ broadening.
On the theoretical front, refinement of the treatment of $k_t$ physics
could take many different forms, from the analog of ``non-perturbative
functions''
in Drell-Yan $p_t$ resummation formalism (cf. previous section) and the
$k_t$-dependent parton distributions associated with the new
``high-energy factorization theorem''
\cite{ktfac}, to the interface of the parton shower calculations and
NLO QCD (as performed
for W production in \cite{baer}).

We clearly need a better understanding of the
underlying physics before low-$p_t$ photon measurements can be confidently
relied upon as the primary source of information on gluons in
a global QCD analysis.  For the time being, the better determination of the
elusive $G(x,Q)$ must depend on a
variety of sources of information including DIS,
direct photon, and differential jet cross-sections.

\section{Acknowledgements}

We thank Werner Vogelsang and Geoff Bodwin for useful
discussions.  This work was supported in part by the
Department of Energy, National Science Foundation, and
the Texas National Research Laboratory Commission.


\end{document}